\documentclass[useAMS,usenatbib]{mnras}

\usepackage{amsmath}
\usepackage{amssymb}
\usepackage{graphicx}
\usepackage{dcolumn}
\usepackage{bm}
\usepackage{ulem}

\title[Plasma screening of nuclear fusion reactions]{Plasma screening of 
nuclear fusion reactions in liquid layers of compact degenerate stars: 
a first-principle study}

\author[D. A. Baiko]
{D. A. Baiko\thanks{E-mail:baiko@astro.ioffe.ru} \\
Ioffe Institute, Politekhnicheskaya 26, 194021 Saint Petersburg, 
Russia}

\begin{document}

\label{firstpage}
\date{Accepted; Received ; in original form}

\pagerange{\pageref{firstpage}--\pageref{lastpage}} \pubyear{2018}

\maketitle

\begin{abstract}
A reliable description of nuclear fusion reactions in inner layers of 
white dwarfs and envelopes of neutron stars is important for realistic 
modelling of a wide range of observable astrophysical phenomena from 
accreting neutron stars to type Ia supernovae. We study the problem of 
screening of the Coulomb barrier impeding the reactions, by a plasma 
surrounding the fusing nuclei. Numerical calculations of the screening 
factor are performed from the first principles with the aid of 
quantum-mechanical path integrals in the model of a one-component 
plasma of atomic nuclei for temperatures and densities typical for 
dense liquid layers of compact degenerate stars. We do not rely on 
various quasiclassic approximations widely used in the literature, 
such as factoring-out the tunneling process, tunneling in an average 
spherically symmetric mean-force potential, usage of classic free 
energies and pair correlation functions, linear mixing rule and so on. 
In general, a good agreement with earlier results from the 
thermonuclear limit to $\Gamma \sim 100$ is found. For a very strongly 
coupled liquid $100 \lesssim \Gamma \leq 175$, a deviation from 
currently used parametrisations of the reaction rates is discovered 
and approximated by a simple analytic expression. The developed method 
of nuclear reaction rate calculations with account of plasma screening 
can be extended to ion mixtures and crystallised phases of stellar 
matter.
\end{abstract}


\section{Introduction}
\label{intro}
Reprocessing of matter accreted onto a compact degenerate star, a white 
dwarf or a neutron star, from a binary companion is associated with a 
number of exciting astrophysical phenomena. The accreted 
material undergoes a series of nuclear transformations via an extensive 
network of reactions involving as many as hundreds of isotopes. In some 
sources, for instance, in soft X-ray transients with transiently 
accreting neutron stars, this `nuclear burning' proceeds in a 
quasisteady-state fashion. The reactions occur throughout the accreted 
crust, slowly alter its composition and properties, determine the 
thermal state of the whole neutron star, and power its quiescent 
emission \citep*[e.g.,][and references therein]{SGC21}. 
In other sources, the reactions may enter a thermal runaway 
regime and produce spectacular explosions. These include type I X-ray 
bursts and superbursts from accreting neutron stars as well as novae 
and, possibly, type Ia supernovae from accreting white dwarfs
\citep*[e.g.,][]{PJS14,Z17,S19}. 

Mergers of compact stars also achieve extreme physical conditions and 
trigger large scale nucleosynthesis. A merger of two white dwarfs can 
be accompanied by a nuclear detonation of the core material which is 
another channel for a type Ia supernova \citep[e.g.,][]{R20}. 
Alternatively, two white dwarfs may merge into a single massive one 
with a distinctive composition \citep*[e.g.,][]{JKS11}. Lastly, when two 
neutron stars merge, one anticipates a major transformation of their 
crustal matter.

In order to model these phenomena and their outcomes realistically, a 
reliable microphysics is required. In particular, an accurate 
description of reactions involved seems to be really important.
One of the key reaction types are nuclear fusion reactions. Let us 
consider an element of an outer neutron star crust or of a white 
dwarf core and assume that there are only fully ionized atoms (i.e. 
atomic nuclei) and electrons present in this matter element. Then, 
a pair of neighbouring nuclei may suffer a quantum mechanical tunneling 
and fuse. A nuclear reaction occurs and new nuclei form. This 
process results in an energy release and a change of matter composition. 

It is well-known that the plasma surrounding the fusing nuclei modifies 
the fusion probability \citep[][]{W40,S48,S54}. Effectively, other ions  
push the pair of fusing nuclei towards each other forcing them to be 
closer and to spend more time near each other thereby enhancing the 
tunneling probability \citep[e.g.,][]{O97}. On top of that, the 
electron density responds to the presence of nuclei and screens the 
Coulomb force between them \citep[e.g.,][]{PC13}. Thus, the ambient 
plasma results in a reduction of the Coulomb barrier and an 
amplification of the reaction rates. These effects are often referred 
to as plasma screening of nuclear reactions 
\citep[for reviews, see][]{YS89,I93}.

The main topic of this paper is the ion contribution to 
plasma screening which, in spite of extensive studies, is still 
subject to some theoretical uncertainty. We shall treat electrons 
simply as a uniform incompressible charge-compensating background.
\citep[Current state of the art in the theory of electron screening is 
summarised in][and references therein.]{PC13} 
Moreover, we shall limit our study to the plasma with nuclei of only 
one sort (with the charge number $Z_i$ and mass $m_i$) deferring a 
more astrophysically relevant case of multi-ionic mixtures to a future 
work \citep[see][for state of the art in this area]{CDW09}. In plasma 
physics, such a model is known as a one-component plasma (OCP).

From a practical standpoint, one is interested in the 
rate of nuclear reactions per unit volume at a given temperature $T$ 
and density $\rho$. It can be written as \citep[e.g.,][]{I93}
\begin{equation}
      R = \frac{n_i^2 a_{\rm B}}{\pi \hbar} S(E_{\rm pk}) g(0)~,
\label{Rdef}
\end{equation}
where $n_i$ is the ion density, $a_{\rm B} = \hbar^2 /(m_i Z_i^2 e^2)$ 
is the ion Bohr radius, $E_{\rm pk}$ is the Gamow-peak energy 
\citep[see also][]{CDW09} and 
$S(E)$ is the astrophysical factor. The latter quantity contains all 
the information about nuclear aspects of the reaction. For non-resonant 
reactions, it is a relatively slowly 
varying function of fusing nuclei energy and, in general, it is not 
known very precisely 
\citep[typical uncertainty is a factor of 2--10, e.g.,][and references 
therein]{Betal10}. We shall not discuss the astrophysical factors in 
this work. 

\begin{figure}                                           
\begin{center}                                              
\leavevmode                                                 
\includegraphics[height=74mm,bb=33 5 378 362,clip]{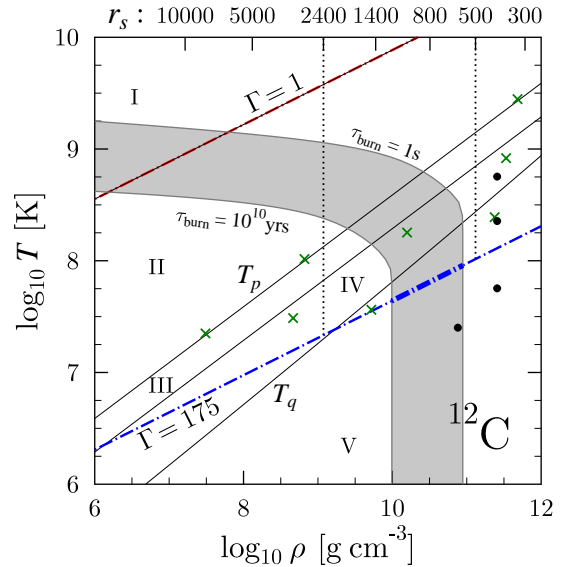} 
\end{center}                                                
\vspace{-0.4cm} 
\caption[]{$T$-$\rho$ plane of fully ionized carbon matter modelled as 
an OCP. Shaded strip is the `relevance strip', the parameter 
range across which the burning time $\tau_{\rm burn}$ changes from 
1 second to $10^{10}$ years. Roman numerals (I--V) indicate various 
plasma screening domains, their boundaries are plotted by thin solid 
lines. Dot-dashed lines show gas-liquid and 
liquid-solid transitions. Crosses and big dots are positions of 
datapoints of \citet{O97} and \citet{MP05}, respectively. Dotted lines 
bound the density range where the present calculations are performed.
Upper $x$-axis is labelled by $r_s$ values. 
}                                             
\label{strip}
\end{figure}

Finally, $g(0)$ is the ion pair correlation function of the plasma at 
zero separation. It is this quantity which encodes the plasma screening 
effect and is the main subject of our 
interest.\footnote{\citet{BS97} and \citet{S10} have 
developed a more general theory of plasma screening, which goes beyond
Eq.\ (\ref{Rdef}). Presently, an application of this theory does not 
look practical, which makes it hard to gauge its actual impact on 
nuclear reaction rates.}
Clearly, if the plasma 
were classic, $g(0)$ would be strictly equal to zero due to the Coulomb 
repulsion of bare nuclei. In reality though, the nuclei can 
tunnel through the Coulomb barrier and there is a non-zero 
(albeit very small) probability for a pair of nuclei to occupy the 
same spatial point, so that the quantum-mechanical $g(0) \ne 0$. If the 
plasma screening were turned off, the  
quantum-mechanical $g(0)$ would be equal to the textbook 
thermonuclear limit \citep[e.g.,][]{C83}. If the screening is weak, 
which is the case in the gaseous ion plasma (domain I in 
Fig.\ \ref{strip}),  
then $g(0)$ can be deduced from the well-known 
Debye-H{\"u}ckel approximation \citep[][]{S54}. These limits of almost
pure thermonuclear fusion are relevant for ordinary stars, outermost 
layers of degenerate stars, and for primordial nucleosynthesis in the 
early Universe.

As one switches to inner layers of compact stars, the screening effect 
becomes much more pronounced. 
For applications, one is interested in an intermediate range of physical 
parameters where the reaction rate $R$ is neither too big nor too small. 
Otherwise, either the assumed composition would be unrealistic and 
heavier nuclei would have to be considered instead, or, by contrast, 
nothing would be happening in the matter element. Since this is a 
well-researched problem, the important parameter range is, in principle, 
already known rather accurately. For carbon plasma, it is shown by a 
shaded strip in Fig.\ \ref{strip}, which is adopted with some 
modifications from Fig.\ 1 of \citet*{CDWY07}. 

The strip spans the parameter range where the burning time, defined as
$\tau_{\rm burn}= n_i/R$, varies from 1 second to $10^{10}$ years. 
The burning time can be phenomenally sensitive to density and/or 
temperature. For instance, in the high-density 
(pycnonuclear) regime in Fig.\ \ref{strip}, it changes by 17 orders of 
magnitude if the density changes by less than an order. This implies 
that any reasonably anticipated improvements of the reaction rates will 
result in only barely visible modifications of this practically 
relevant physical domain (for brevity, we shall refer to this domain 
as the `relevance strip').

%

The fact that the strip is shown specifically for carbon plasma does 
not limit the discussion because the problem for other ions 
(as long as it is still the OCP model and the nuclear reaction in 
question is non-resonant) is self-similar and can be 
solved by simple scaling with the aid of two dimensionless parameters. 
These are the familiar Coulomb coupling parameter 
$\Gamma = Z_i^2 e^2/(a_i T)$ and the ion density  
parameter $r_s =a_i/a_{\rm B}$. In this case, 
$a_i =(4 \pi n_i/3)^{-1/3}$ is the Wigner-Seitz radius. Let us remind 
that $\Gamma = 1$ corresponds to the liquid-gas continuous transition 
and $\Gamma = 175$ is the melting phase transition for the classic OCP 
\citep[e.g.,][]{HPY07}. These equations are represented by dot-dashed 
lines in Fig.\ \ref{strip}.  

Since the pioneering work of \citet{SVH69}, it has been 
customary to divide the temperature-density plane into domains
corresponding to different plasma screening physics. These are also
shown in Fig.\ \ref{strip} and marked by roman numerals (I--V). Their
boundaries \citep[as given in][]{CDWY07} are displayed by thin solid 
lines described by the following equations: $\Gamma=1$, $T=T_p$, 
$T=0.5 T_p$, and $T=T_q$, where 
$T_p = \hbar \sqrt{4 \pi n_i Z^2 e^2 / m_i}$
is the ion plasma temperature and 
$T_q = T_p/ \ln{(r_s/3)}$.
The relevance strip crosses these domains and lines separating gas, 
liquid, and solid phases in such a way that, in the gas phase, the strip 
is always in domain I (`thermonuclear with weak screening'). In the 
solid phase, it is always in domain V (`$T=0$ pycnonuclear'). 
In the liquid phase, the strip traverses domain II (`thermonuclear 
with strong screening'), intermediate domains III and IV, and domain V. 

Thus, the segment of the melting line belonging to the relevance 
strip (shown by a thicker dot-dashed line in Fig.\ \ref{strip}) is 
completely within the $T=0$ pycnonuclear domain V in which the 
reaction rates are supposedly independent of temperature. This is 
actually surprising, because melting/crystallisation in such a plasma 
is certainly a thermal phenomenon. Hence, one tends to expect the 
temperature dependence of the rates in the liquid to continue 
all the way down to crystallisation and, to some extent, into the 
solid phase as well. It is worth  
to keep this observation in mind when we analyse various 
results in Sections \ref{formal} and \ref{results}.   

In general, the domain boundaries are deduced from qualitative arguments
and should not be taken too literally. For instance, one can find three
versions of the upper boundary of the $T=0$ pycnonuclear domain in
\citet{SVH69}: in their Fig.\ 1, p.\ 187, and p.\ 198. For carbon at
$\rho = 3 \times 10^{10}$ g cm$^{-3}$ ($r_s \approx 818$), they become
$3.7 \times 10^6$ K, $7 \times 10^6$ K, and $1.2 \times 10^8$ K, 
respectively. The last estimate is virtually equal to $T_q$. 

In this work, we shall focus on the liquid and gas plasma phase 
$\Gamma \leq 175$. Particular attention will be paid to the very
strongly coupled liquid at $100 \lesssim \Gamma \leq 175$. In terms of 
density, we shall limit ourselves to the range 
$500 \leq r_s \leq 2400$. For carbon plasma, this translates into 
$11.12 \geq \log_{10}{\rho} \geq 9.07$. In Fig.\ \ref{strip}, the 
physical range to be studied is confined by two vertical 
dotted lines and the melting line. 

The range of temperatures and densities addressed in this work 
covers completely the areas of the relevance strip belonging to 
domains III, IV, and V (in the liquid phase) and partially covers its 
portion belonging to domain II. The reaction rates in domains I and II 
are generally considered known reliably (this will be further confirmed 
by our results). By contrast, in domains III, IV, and V, the rates are 
subject to some uncertainties \citep[e.g.,][]{CDWY07}. Thus, if we 
manage to calculate the reaction rates in our target temperature and 
density range, the plasma screening problem in the liquid phase of the 
OCP will be fully solved [within the framework of Eq.\ (\ref{Rdef})].

\section{Previous work}
Following the original work of \citet{W40}, \citet{S48}, and 
\citet{S54}, plasma screening of nuclear reactions has been 
investigated in many excellent papers 
\citep[see, e.g.,][for an extensive compilation of these works]{C21}. 
Broadly speaking, they can be divided into 3 groups: 
$(i)$ earlier research, oftentimes semi-analytic, culminating in the 
paper by \citet{AJ78}; $(ii)$ a series of works based on more advanced 
classic Monte Carlo (MC) simulations including \citet*{IKM90}, 
\citet*{OII91}, \citet{CDWY07}; $(iii)$ quantum MC simulations of 
\citet{O97} and \citet{MP05}. For our purposes here, the most 
important are the quantum calculations of the last group as well as the 
most recent classic MC based study by \citet{CDWY07}.

The present paper is devoted to a first-principle calculation of true 
quantum-mechanical $g(0)$ by the method of path-integral Monte Carlo 
(PIMC) simulations. Currently, PIMC is the preferred method for 
solving quantum many-body problems of this sort. For the first time, 
this task has been addressed by \citet{O97}, who expressed the ratio of 
the true $g(0)$ to its thermonuclear limit via the ratios of ensemble 
averages of certain Boltzmann exponentials. These averages were 
evaluated directly with the aid of MC sampling. The range of physical 
parameters chosen by Ogata intersected the relevance strip. The 
positions of some of his datapoints are shown by crosses in 
Fig.\ \ref{strip}. 
For practical applications, \citet{O97} approximated numerical results 
by a convenient analytic formula which allows for an easy comparison 
with any new results. Ogata has shown that it was sufficient to consider
distinguishable particles, i.e. the ion statistics played no role.
It appears \citep[cf.\ Fig.\ 4 of][]{O97} that, when sampling 
exponentials, he had to deal with rather strongly fluctuating 
quantities.      

The only other PIMC calculation of $g(0)$ has been reported by 
\citet{MP05}. It is not 100\% clear what method of extracting $g(0)$ 
from PIMC simulations was used by these authors. It appears that they 
extrapolated the numerical values of $\ln{[g(r)]}$ from finite 
$r > a_i$ which could be sampled in MC runs (close encounters with 
$r < a_i$ are exponentially rare due to the Coulomb barrier) to 
$r \to 0$. Unfortunately, the results of \citet{MP05} in the liquid 
were obtained outside of the relevance strip, at such conditions where 
$g(0)$ was relatively big and the burning time was extremely brief, 
much shorter than 1 second. Four of their points are shown by dots in 
Fig.\ \ref{strip} while the other 32 points lie outside of the 
temperature or density ranges spanned by this figure. The authors 
proposed no way of extrapolating the results into the physically 
relevant domain and, for this reason, their data cannot be directly 
compared with ours.
           
A totally different method was employed by \citet{CDWY07}. These 
authors worked with a pair correlation function found in {\it classic} 
MC simulations. Classic $g(0)=0$. However, one can define the 
effective spherically symmetric pair potential $U(r)$ according to
$g(r) = \exp{[-U(r)/T]}$. This potential contains a divergent Coulomb 
term and a finite term responsible for screening. The latter was 
extrapolated to $r \to 0$. Once again, close encounters are 
extremely rare, thus one has to base the extrapolation on numerical 
results at $r > a_i$. After that, the tunneling probability and 
respective $g(0)$ were found via the standard quantum-mechanical WKB 
technique. The same method was used earlier by \citet{IKM90} with 
similar results. In \citet{CDWY07}, it was stated that their 
extrapolated potential was more advanced than that of \citet{IKM90}. 
\citet{CDWY07} also approximated their numerical results by a rather 
convenient analytic formula. We note in passing, that in Line (D) 
of their Tab. III, when describing the analytic formula of \citet{O97}, 
\citet{CDWY07} omitted a term given by Eq.\ (22) of \citet{O97}.    

One may wonder, whether calculations based on classic MC have a chance 
of being accurate? In a related problem, calculations of the 
liquid OCP energy by the PIMC method \citep[][]{JC96,B19} 
reveal a significant difference with the classic MC energy in 
the density range $600 \leq r_s \leq 2400$ relevant for the present 
work. The discrepancy becomes worse as the system becomes more quantum. 
Obviously though, the work of \citet{CDWY07} is not just classic but 
takes into account the crucial quantum tunneling effect. 

Conversely, one may ask why use an extremely time-consuming PIMC method 
given the relative ease of classic MC based or semi-analytic treatments?   
To answer this question, one has to realise that the accuracy of an 
approach such as that of \citet{CDWY07} is, in fact, unknown until it 
is compared to a first-principle calculation as a number of 
effects remains unaccounted for. First of all, not a single 
tunneling event occurs in a static spherically-symmetric 
potential. The actual potential, produced by many ions, fluctuates in 
time and space. Even its spherically-symmetric average is likely 
affected by the quantum nature of surrounding particles and differs from
the respective average deduced from classic MC. Each tunneling 
is a three-dimensional anisotropic process which, in general, cannot 
be described by a one-dimensional radial WKB formula. Finally, it is not
clear whether the extrapolation procedure for $U(r)$ is unambiguous.

\citet{CDWY07} claim that none of these effects is 
expected to be strong in thermonuclear regime with strong plasma 
screening (domain II). In domains III, IV, and V, the situation is less 
optimistic. PIMC allows one to take all these effects into account, 
provide reliable numerical results in the most problematic parameter 
range, and quantify the precision of the quasiclassic treatment.

\section{Formalism}
\label{formal}
\subsection{General formula}
The pair correlation function at a spatial separation ${\bm r}$ reads
\citep[e.g.,][]{O97}
\begin{equation}
    g({\bm r}) = \frac{\Omega \, {\rm Sp} \left[\delta({\bm r}_{12} 
    - {\bm r}) e^{-\beta {\cal H}}\right]}{{\rm Sp} 
    \left[ e^{-\beta {\cal H}} \right]}~,
\label{grdef}
\end{equation}
where $\Omega$ is the volume, 
`Sp' means trace,
${\cal H}$ is the system Hamiltonian, 
$\beta = 1/T$, ${\bm r}_{12} = {\bm r}_{1} - {\bm r}_{2}$, ${\bm r}_j$
is the coordinate of the $j$-th particle (i.e. a wave function 
argument over which one integrates when computing matrix elements),
$j=1, 2, \ldots N$. Finally, $N$ is the total number of particles. 

The traces in Eq.\ (\ref{grdef}) can be evaluated in any 
representation but we shall use the coordinate representation where the 
quantum numbers are particle coordinates 
${\bm R}_1$, ${\bm R}_2$, \ldots ${\bm R}_N$ (over which one integrates 
when computing trace). A calligraphic 
${\cal R}= \left\{ {\bm R}_1, {\bm R}_2, \ldots {\bm R}_N \right\}$
denotes a full set of these quantum numbers specifying a basis function 
of the coordinate representation.

\begin{figure}                                           
\begin{center}                                              
\leavevmode                                                 
\includegraphics[height=74mm,bb=33 12 372 347,clip]{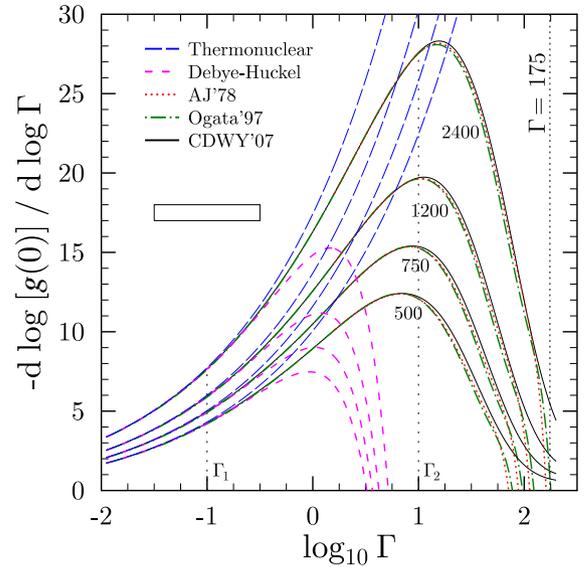} 
\end{center}                                                
\vspace{-0.4cm} 
\caption[]{Minus the logarithmic derivative of $g(0)$ with respect to 
$\log{\Gamma}$ in the thermonuclear limit (long-dashed), 
Debye-H{\"u}ckel approximation (short-dashed), models of \citet{AJ78} 
(dotted), \citet{O97} (dot-dashed), and \citet{CDWY07} (solid). The 
rectangle demonstrates an area under a curve producing a 10-fold change 
of $g(0)$ and the respective reaction rates. Numbers near the curves 
are $r_s$ values.
}                                             
\label{integrand}
\end{figure}
%

In the denominator of Eq.\ (\ref{grdef}), one has the partition 
function and the numerator can be expected to behave in a similar 
way. In simulations, one typically tries to avoid direct computation of 
the partition function because it involves averaging exponentials which 
produces excessive fluctuations during sampling \citep[e.g.,][]{C95}. 
Instead, one computes energy, which is 
a much better behaved quantity, and then integrates it over temperature
to obtain the Helmholtz free energy (proportional to the logarithm of 
the partition function). 

Acting in the same spirit let us consider
\begin{equation}
      \frac{{\rm d}}{{\rm d} \beta} \ln{g({\bm r})} = 
      \frac{ {\rm Sp} \left[{\cal H} e^{-\beta {\cal H}}\right]}
           { {\rm Sp} \left[e^{-\beta {\cal H}} \right]} -
      \frac{ {\rm Sp} \left[\delta({\bm r}_{12}-{\bm r}) {\cal H} 
            e^{-\beta {\cal H}}\right]}
           { {\rm Sp} \left[\delta({\bm r}_{12}-{\bm r}) 
           e^{-\beta {\cal H}} \right]}~. 
\label{ddbeta}
\end{equation}
The method that we outline allows one to find $g({\bm r})$ at any 
${\bm r}$, but from now on we shall focus on $g(0)$. Consequently, we 
obtain
\begin{equation}
           \ln{\frac{g(0,\Gamma_2)}{g(0,\Gamma_1)}} =
              \int^{\ln{\Gamma_2}}_{\ln{\Gamma_1}} 
            {\rm d} \ln{\Gamma} \,\,   
            \beta \big( \langle{\cal H} \rangle - \langle{\cal H} 
            \rangle_{{\bm r}_{12}=0} \big)~.          
\label{integral}
\end{equation}
In this case, $\langle{\cal H} \rangle$, given by the first fraction on 
the right-hand side of Eq.\ (\ref{ddbeta}), is the ordinary energy
of the system, and 
\begin{equation}
       \langle{\cal H} \rangle_{{\bm r}_{12}={\bm r}} =
       \frac{ {\rm Sp} \left[\delta({\bm r}_{12}-{\bm r}) {\cal H} 
            e^{-\beta {\cal H}}\right]}
           { {\rm Sp} \left[\delta({\bm r}_{12}-{\bm r}) 
           e^{-\beta {\cal H}} \right]} 
\label{H0def}
\end{equation}
is a similar average restricted to functions with 
${\bm R}_1 = {\bm R}_2+{\bm r}$. For clarity, we included $\Gamma$
into the list of arguments of the pair correlation function in 
Eq.\ (\ref{integral}). Thus, if the quantity   
$\beta ( \langle{\cal H} \rangle - \langle{\cal H} 
\rangle_{{\bm r}_{12}=0})$ is calculated for all $\Gamma$ greater than
some fixed value $\Gamma_1$, it can be 
integrated over $\ln{\Gamma}$ to find $g(0,\Gamma_2)$ at any 
$\Gamma_2$ provided $g(0,\Gamma_1)$ is known.

\subsection{Illustration}
To get a better feel of this quantity, it is helpful to plot
the integrands generating $g(0)$ via Eq.\ (\ref{integral}) in already 
existing models. In Fig.\ \ref{integrand}, curves of 
different types correspond to different models at $r_s$ values 
indicated next to them. The area under any curve between any two 
$\Gamma$, say $\Gamma_1=0.1$ and $\Gamma_2=10$ is equal to minus 
decimal logarithm of the factor by which $g(0)$ drops from 
$\Gamma_1$ to $\Gamma_2$ in the respective model. For illustration, 
the rectangle shows an area changing the decimal logarithm by 1. We 
see, for instance, that the area under the $r_s=2400$ curves is 
much larger than that under the $r_s=500$ curves, which means that 
$g(0)$ and reaction rates at $r_s=2400$ drop much faster than those 
at $r_s=500$ over the same $\Gamma$-range. 

Long-dashed and short-dashed curves display 
the thermonuclear and Debye-H{\"u}ckel limits, respectively. The main 
qualitative difference between them is the appearance of maxima at 
$\Gamma \sim 1$ and descending portions of the curves in the 
Debye-H{\"u}ckel model. These features represent the 
main effect of the ion screening. They prevent an extremely rapid drop 
of $g(0)$ with increase of $\Gamma$ predicted by the pure thermonuclear
curves. 

However, a comparison with other, more sophisticated, models 
demonstrates that the importance of ion screening is greatly 
overestimated by the Debye-H{\"u}ckel approximation in the liquid 
regime $\Gamma > 1$. Dotted, dot-dashed and solid curves show the 
results of \citet{AJ78}, \citet{O97}, and \citet{CDWY07}, respectively, 
and the areas under these curves are obviously bigger than those under 
the short-dashed curves. Accordingly, the decrease of $g(0)$ with 
$\Gamma$ obtained by these authors is much stronger than in the 
Debye-H{\"u}ckel model. The three parametrisations agree with each 
other relatively well. We do observe some differences between the 
descending segments of the curves at $\Gamma \gtrsim 10$. 
Most apparent differences occur at 
$\Gamma \gtrsim 100$ where the curves of \citet{AJ78} and \citet{O97}
drop to zero. This means that at some point in the {\it liquid} phase 
their $g(0)$ stop decreasing with decrease of temperature so that
the reaction rates become temperature independent. This corresponds
to the onset of the pycnonuclear regime V (cf.\ Section \ref{intro})
and is qualitatively consistent with the $T_q$ value above the melting 
temperature. On the contrary, $g(0)$ as predicted by \citet{CDWY07} 
continues to decrease (i.e. depends on temperature) all the way down to 
the crystallisation line $\Gamma = 175$.       

At low $\Gamma$, all models merge which means that $g(0,\Gamma_1)$ can 
be taken from the thermonuclear limit at a sufficiently low $\Gamma_1$.

\subsection{PIMC implementation}
The energy of a quantum Coulomb liquid $\langle{\cal H} \rangle$ has 
been recently calculated by the PIMC
method \citep[][]{B19} and we refer to this work for details 
of our specific realisation of the method \citep[see also][]{C95,JC96}. 
For both, $\langle{\cal H} \rangle$ and 
$\langle{\cal H} \rangle_{{\bm r}_{12}=0}$, we split 
$e^{-\beta {\cal H}} = \left(e^{-\tau {\cal H}}\right)^M$, 
where $\tau = \beta/M$, $M$ is a positive integer. In the numerators of
Eq.\ (\ref{ddbeta}), 
one can commute ${\cal H}$ with an arbitrary number of factors 
$e^{-\tau {\cal H}}$ which should not affect the result. In what 
follows, this is referred to as a slice number (or a bead number) 
selection for an energy estimate. 

In view of the Trotter formula 
\begin{equation}
          e^{-\beta ({\cal T} + {\cal V})} = \lim_{M\to \infty}
          \left[e^{-\tau {\cal T}}  e^{-\tau {\cal V}} \right]^M
\label{trotter}
\end{equation}
(where ${\cal T}$ and ${\cal V}$ are kinetic and potential energy 
operators, respectively), one can approximately replace 
$e^{-\tau {\cal H}} \to e^{-\tau {\cal T}} e^{-\tau {\cal V}}$ 
everywhere. This is known as the primitive approximation (its accuracy 
should be improving with decrease of $\tau$ and is further discussed 
below). 

Then one can insert full sets of basis
functions ${\cal R}_l = \left\{ {\bm R}^{(l)}_1, {\bm R}^{(l)}_2, \ldots 
{\bm R}^{(l)}_N \right\}$ between all operators in Eq.\ (\ref{ddbeta}) 
($l$ enumerates basis sets). Any matrix elements of 
$e^{-\tau {\cal T}}$ and $e^{-\tau {\cal V}}$ can be easily evaluated
analytically in the coordinate representation \citep[e.g.,][]{C95}.
For instance, 
\begin{eqnarray}
        && \int {\rm d}{\cal R} \langle {\cal R}_1 \vert 
        e^{-\tau {\cal T}} \vert {\cal R}  \rangle
        \langle {\cal R} \vert e^{-\tau {\cal V}} \vert {\cal R}_2 
        \rangle 
\nonumber \\        
        &=&  \int   
           \frac{{\rm d}{\cal R}}{(4 \pi \lambda \tau)^{3N/2}} 
        \exp{\left[-\frac{({\cal R}_1-{\cal R})^2}{4 \lambda \tau} 
        \right] e^{-\tau V({\cal R})} \delta ({\cal R}-{\cal R}_2)}
\nonumber \\
       &=&    
                \frac{1}{(4 \pi \lambda \tau)^{3N/2}} 
        \exp{\left[-\frac{({\cal R}_1-{\cal R}_2)^2}{4 \lambda \tau} 
        -\tau V({\cal R}_2) \right]}~,
\end{eqnarray}
where $\lambda = \hbar^2/(2m_i)$ and $V({\cal R})$ is the system 
potential energy when ion coordinates are equal to ${\cal R}$. 
A matrix element of the Hamiltonian between these basis functions in 
Eq.\ (\ref{ddbeta}) can be also readily derived which yields an 
expression called an {\it energy estimator}.

After some algebra, one is left with $M+1$ basis functions ${\cal R}_0$, 
${\cal R}_1$, \ldots ${\cal R}_M$, ${\cal R}_0 = {\cal R}_M$ being 
final and initial states equal to each other because we calculate 
trace. Functions ${\cal R}_1$, ${\cal R}_2$, \ldots ${\cal R}_{M-1}$ 
describe intermediate states.  We denote the energy estimator as 
$H({\cal R}_m)$ where $m$ (which can take values 1, 2, \ldots $M-1$) 
is the slice number selected for the energy estimate. A lengthy 
explicit formula for $H({\cal R}_m)$ can be found, e.g., in Eq.\ (6.5) 
of \citet{C95}.      

Combining the above formulae, one obtains 
\begin{eqnarray}
        \langle{\cal H} \rangle &=& \int {\rm d}\sigma \pi(\sigma) 
        H({\cal R}_m)~,
\label{work_eq}\\
        \langle{\cal H} \rangle_{{\bm r}_{12}=0} &=& 
        \int {\rm d}\sigma_0 \pi_0(\sigma_0) H({\cal R}_m)~, 
\nonumber\\
        \pi(\sigma) &=& \frac{1}{Z} 
        \exp{\left[ -\sum_{l=1}^M S_l \right]}~,
\nonumber\\
           e^{-S_l} &=& \frac{1}{(4 \pi \lambda \tau)^{3N/2}}
      \exp{\left[- 
      \frac{({\cal R}_l-{\cal R}_{l-1})^2}{4 \lambda \tau} 
      - \tau V({\cal R}_l) \right]}
\nonumber\\
       \sigma &=& \left\{ {\cal R}_1, {\cal R}_2 \ldots, 
       {\cal R}_M = {\cal R}_0 \right\}~.
\nonumber                                   
\end{eqnarray}
In this case, $\sigma$ is the $3NM$-dimensional domain of quantum 
numbers (coordinates of $N$ ions in all $M$ states) and $\pi(\sigma)$
is a product of matrix elements of the operator exponentials. Since 
$\pi(\sigma)$ is strictly positive, while $Z$ [coming from the 
denominator in Eq.\ (\ref{ddbeta})] ensures that $\pi(\sigma)$ 
is normalized to 1, it is natural to interpret $\pi(\sigma)$ as a 
probability distribution, and the task of finding 
$\langle{\cal H} \rangle$ reduces to averaging the energy estimator 
with it. 

The $3(NM-1)$-dimensional domain $\sigma_0$ differs from 
$\sigma$ in that coordinates of particles 1 and 2 in the initial and 
final states always coincide: ${\bm R}^{(0)}_1 = {\bm R}^{(0)}_2$. The 
distribution $\pi_0(\sigma_0)$ is then obtained from $\pi(\sigma)$ by a 
straightforward modification of the kinetic energy terms in 
$e^{-S_1}$ and $e^{-S_M}$. In the potential energy, the singular 
exponential does not enter $\pi_0$ because it is simply a common 
factor (for any finite ${\bm r}$, it is 
$\sim e^{-\tau Z_i^2e^2/|{\bm r}|}$) which cancels out with the 
denominator. 

As is well known \citep[e.g.,][]{C95}, the quantum system at finite $M$
is isomorphic to a system of $N$ classic ring polymers each with $M$ 
beads. The bead coordinates of the $j$-th ring are of course
${\bm R}^{(0)}_j$, ${\bm R}^{(1)}_j$, \ldots 
${\bm R}^{(M)}_j = {\bm R}^{(0)}_j$. In order to find the 
average energy $\langle{\cal H} \rangle$, one studies a system of $N$ 
independent polymers. To calculate 
$\langle{\cal H} \rangle_{{\bm r}_{12}=0}$, the ring polymers 
describing particles 1 and 2 have a common zeroth bead of double mass 
and double charge.  

Sampling is done with the aid of the Metropolis 
algorithm. We attempt two move types, single bead moves and whole 
polymer moves. The details can be found in \citet{B19}. 

Since we are now interested in a relatively small difference of two 
close quantities:
$\langle{\cal H} \rangle - \langle{\cal H}\rangle_{{\bm r}_{12}=0}$,
where each of the averages can be extracted from simulations with 
appreciable errors, it is desirable to work in such a way that would 
result in a cancellation of common sources of variance. In a sense, it 
is the same principle as that governing differential amplifiers very 
popular in electrotechnics: common mode noises cancel but the difference 
between two inputs is amplified. In order to achieve this, we run 
simulations for $\langle{\cal H} \rangle$ and 
$\langle{\cal H}\rangle_{{\bm r}_{12}=0}$ in parallel with identical 
random sequences and the same initial conditions, except that, for 
$\langle{\cal H}\rangle_{{\bm r}_{12}=0}$, we find two particles in the 
initial file with the smallest separation between them, shift the 
particles towards the mid-point positioning them very close to each 
other and merge their zeroth beads right at the mid-point. Furthermore, 
we use the same slice $m$ for energy estimates, same estimator (see 
below) and perform the estimates after the same number of attempted 
moves of the Metropolis process.

Let us add a few more words regarding the slice number $m$ and 
estimator selection for the energy estimates. 
For $\langle{\cal H}\rangle_{{\bm r}_{12}=0}$, the {\it thermodynamic} 
estimator [e.g., Eq.\ (6.7) of \citet{C95}] shows very pronounced 
slice-number dependence, because it relies on equivalence of beads. 
But they are no longer equivalent in the 
$\langle{\cal H} \rangle_{{\bm r}_{12}=0}$ simulation being closer or 
further away from the common bead. Simply put, the thermodynamic 
estimator is not appropriate here. 

Instead, one should use the {\it direct} or {\it Hamiltonian} energy 
estimator [e.g., Eq.\ (6.5) of \citet{C95}], which is just the matrix 
element $H({\cal R}_m)$ introduced above. This estimator turns out to 
be much closer to being $m$-independent. It has an extended 
$m$-independent `shelf' at intermediate $m$, yet, there is a 
noticeable $m$-dependence for several slices near the common bead. 
The reason for this (unwanted) $m$-dependence of the direct estimator 
for $\langle{\cal H} \rangle_{{\bm r}_{12}=0}$ is related to the fact 
that the accuracy of the primitive approximation
$e^{-\tau {\cal H}} \to e^{-\tau {\cal T}} e^{-\tau {\cal V}}$ is not
uniform across slices. It can be expected to be worse for $m$ near the 
common bead because ${\cal V}$ is relatively large there.

An analysis shows that, if one compares, for instance, $m=2$ for a 
particular $\tau$ and $m=4$ for a two times smaller $\tau$ (i.e. two 
times greater $M$), the energy estimate will be closer to the shelf in 
the second case. In other words, with increase of $M$, the shelf widens 
(in imaginary time) towards the common bead. As far as 
$\langle{\cal H} \rangle$ is concerned, there is no dependence on 
$m$ regardless of 
which energy estimator (direct or thermodynamic) is used and both 
estimators, within statistical errors, yield the same result. This 
leads us to a procedure where for 
$\langle{\cal H}\rangle_{{\bm r}_{12}=0}$ we employ the direct 
estimator with $m$ taken at least 5--6 beads away from the common bead, 
i.e. on the shelf, and the same $m$ and the estimator are chosen for 
$\langle{\cal H} \rangle$.      

\begin{figure}                                           
\begin{center}                                              
\leavevmode                                                 
\includegraphics[height=74mm,bb=33 23 373 348,clip]{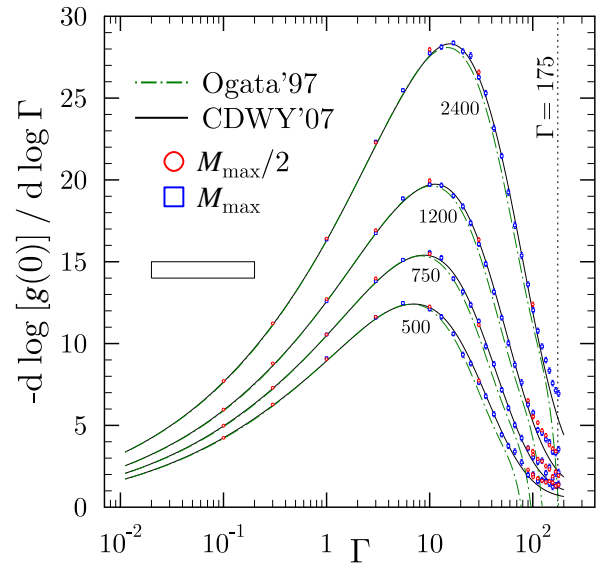} 
\end{center}                                                
\vspace{-0.4cm} 
\caption[]{Simulation results for the integrand of 
Eq.\ (\ref{integral}) for a maximum number of beads (squares) and for
two times fewer beads (circles). The other curves are the same as in 
Fig.\ \ref{integrand}.
}                                             
\label{num_res}
\end{figure}

The primitive approximation becomes exact as $M \to \infty$. In 
practice, one can perform calculations at several $M$ and then 
extrapolate to $1/M \to 0$.  Such an approach is used for calculation 
of liquid and crystal energies 
\citep[][Baiko \& Chugunov, submitted]{B19}. In the present 
problem though, the difference 
$\langle{\cal H} \rangle - \langle{\cal H} \rangle_{{\bm r}_{12}=0}$
has no observable $M$-dependence for sufficiently large $M$
(cf.\ Figs.\ \ref{num_res} and \ref{M-dep}). This is 
likely related to a good cancellation of $M$-dependent terms in the two
averages. The same effect has been also observed earlier by \citet{O97}. 

%

Any first principle simulation is performed with a finite number of 
particles ($N=250$ in our case) and there is always the question 
whether the results are sensitive to $N$ and what would they be in the
thermodynamic limit? The only way to conclusively answer this question
is to perform simulations at several $N$ and analyse the ensuing 
$N$-dependence. This program requires extra resources
and has not been implemented within the framework of this paper. We 
note, however, that \citet{O97} has reported calculations at two values 
of $N=50$ and 100 and found no substantial $N$-dependence.

\section{Results}
\label{results}
In Fig.\ \ref{num_res}, we present results of our simulations (as 
one-sigma bars) for $r_s=500$, 750, 1200 and 2400, along with the 
integrands derived from the analytic fits of \citet{O97} (dot-dashed) 
and \citet{CDWY07} (solid). These curves have already been displayed 
in Fig.\ \ref{integrand}. Squares correspond to a maximum number of 
beads in a ring polymer achieved in simulations, circles are 
for two times fewer beads. 
 
\begin{figure}                                           
\begin{center}                                              
\leavevmode                                                 
\includegraphics[height=74mm,bb=33 27 372 347,clip]{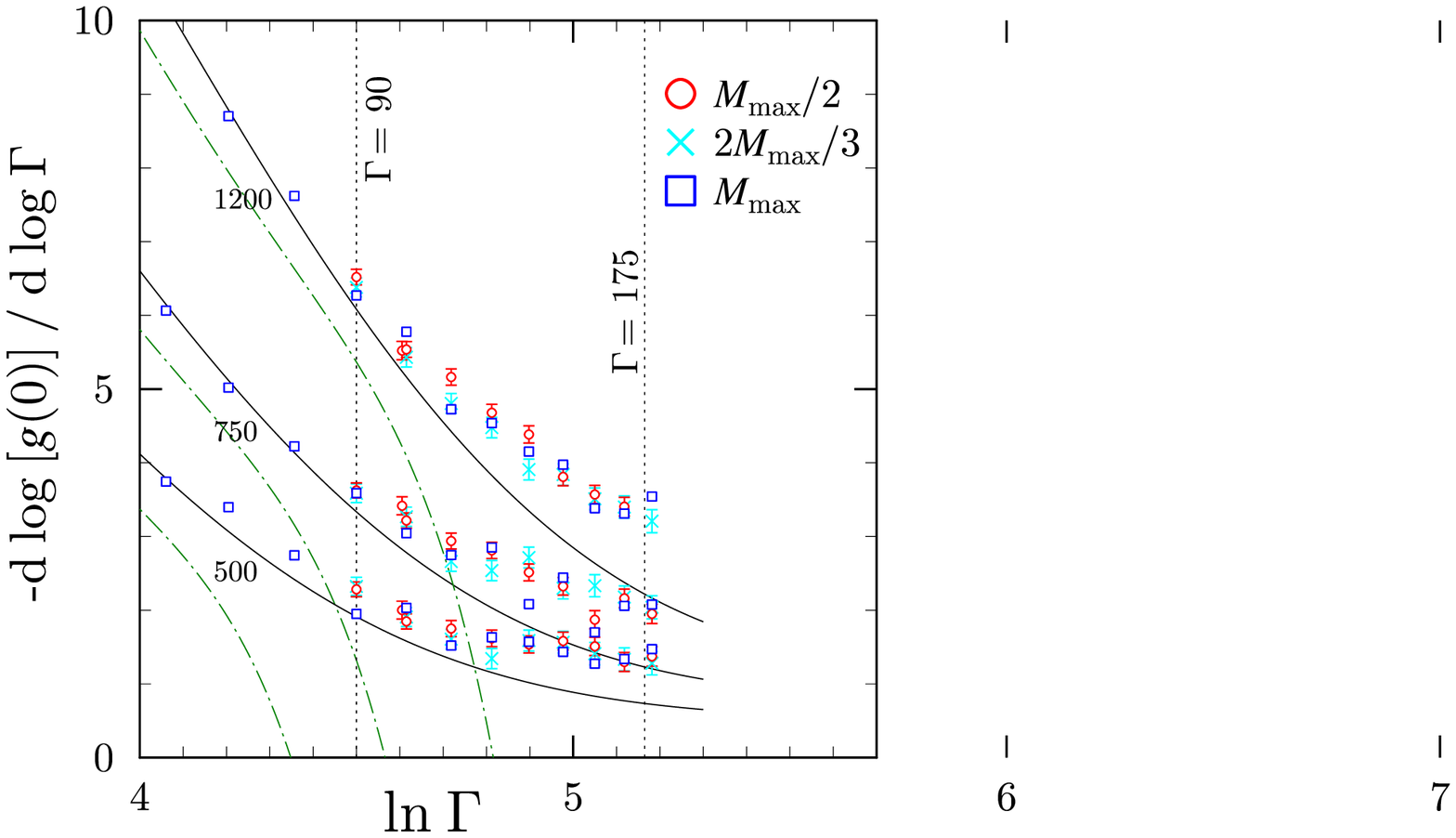} 
\end{center}                                                
\vspace{-0.4cm} 
\caption[]{Zoomed-in high-$\Gamma$ area of Fig.\ \ref{num_res}. Also
shown are calculations at an intermediate number of beads (crosses).
}                                             
\label{M-dep}
\end{figure}
%

We would like to emphasize that the methods and calculations, compared 
in Fig.\ \ref{num_res}, are completely independent. In our approach, 
there is no reference to classic ion 
plasma, its pair correlation function or free energy. Tunneling is not 
treated as a separate, factored-out process. Nevertheless,  
the excellent agreement from the thermonuclear limit up to 
$\Gamma \sim 10$ mutually validates different theoretical approaches.
The thermonuclear limit itself is verified by our numerical results at 
low $\Gamma$.
  
\begin{figure}                                           
\begin{center}                                              
\leavevmode                                                 
\includegraphics[height=74mm,bb=44 31 372 343,clip]{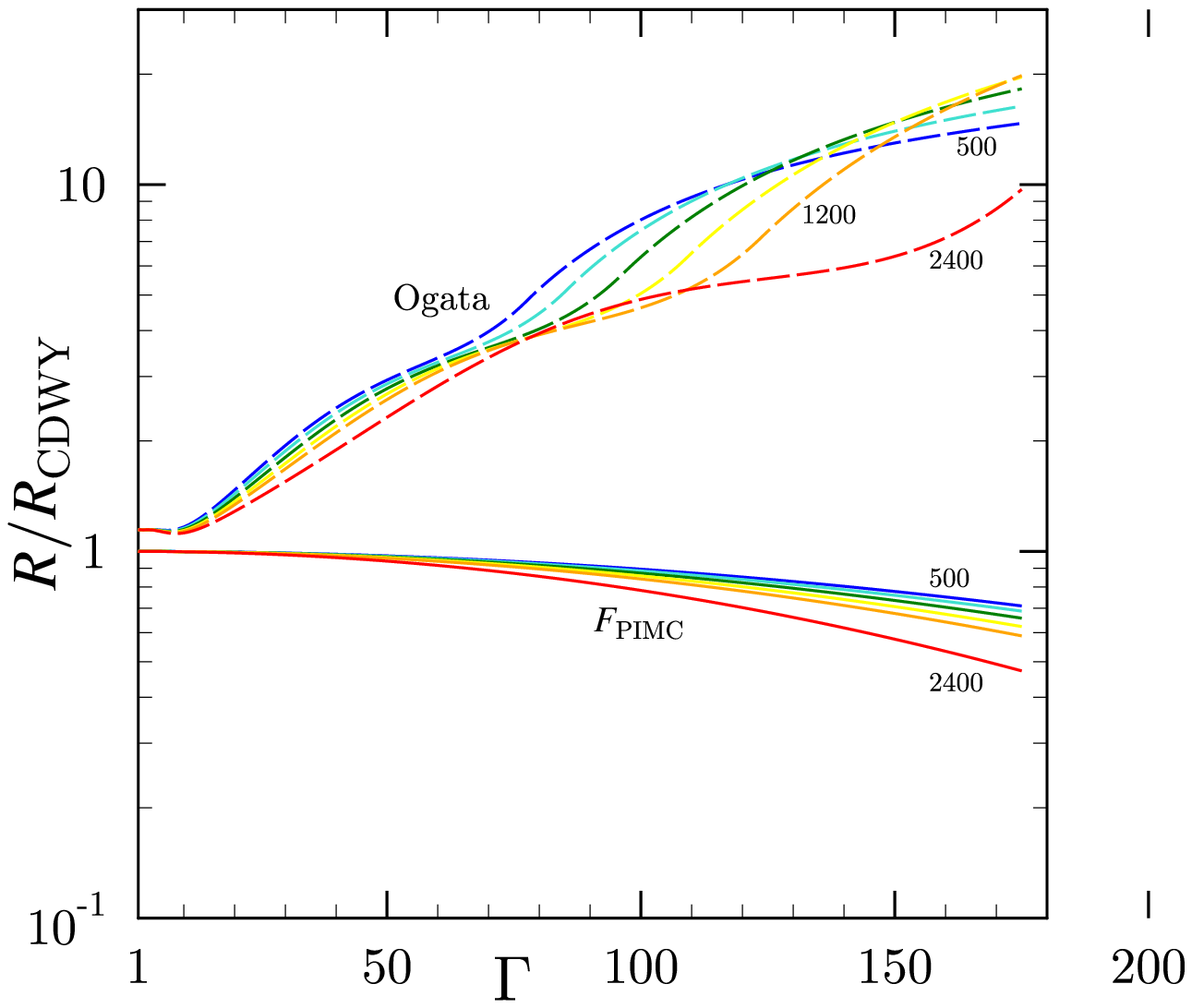} 
\end{center}                                                
\vspace{-0.4cm} 
\caption[]{The ratios of the reaction rates, $R$, as reported by 
\citet{O97} (dashes) and calculated in the present work (solid curves) 
to those due to \citet{CDWY07}, denoted as $R_{\rm CDWY}$. 
Numbers near the curves are the respective $r_s$ values.
}                                             
\label{rates}
\end{figure}

On the descending portions of the curves our data favor the
parametrisation of \citet{CDWY07}. However, at $\Gamma \gtrsim 100$, our
points deviate from their curves and predict somewhat 
smaller $g(0)$ and slower reaction rates (cf.\ Fig.\ \ref{rates}).
Importantly, our data confirm that the integrands do not drop to 
zero in the liquid phase and the reaction rates continue to depend on 
$T$ at least down to the crystallisation (cf.\ Fig.\ \ref{M-dep}).     

Let us note, that different $r_s$ and $\Gamma$ points
have been calculated by PIMC using different random sequences, initial 
conditions, numbers of MC steps, slice numbers at which the energy was
estimated. This explains a higher variance {\it between} our datapoints
than may have been expected based on the error bar sizes characterising
individual MC runs. 

A comparison of squares and circles does not reveal any systematic 
dependence of our results on $M$. A few misses at various $r_s$ and 
$\Gamma$ should be attributed to fluctuations. This point is 
illustrated further in Fig.\ \ref{M-dep} where we consider the 
$\Gamma \gtrsim 100$ region in more detail. Solid curves represent 
analytic fits of \citet{CDWY07}. Symbols show simulation results at 
three different $M$ whose inverse values are evenly spaced. Squares and 
circles are the same data as in Fig.\ \ref{num_res}. In the case of 
squares, we omit the error bars to improve the graph 
readability. They are of about the same size as the drawn bars. Crosses 
correspond to an intermediate number of beads. Basically, this figure 
demonstrates that the $M$-dependence has already saturated in the 
considered $M$-range, and the results at the maximum number of 
beads can be used as the `production' datapoints. Dot-dashed 
curves are fits of \citet{O97} which are clearly in a serious 
disagreement with our calculations in this $\Gamma$-range.

\begin{figure}                                           
\begin{center}                                              
\leavevmode                                                 
\includegraphics[height=74mm,bb=34 28 371 347,clip]{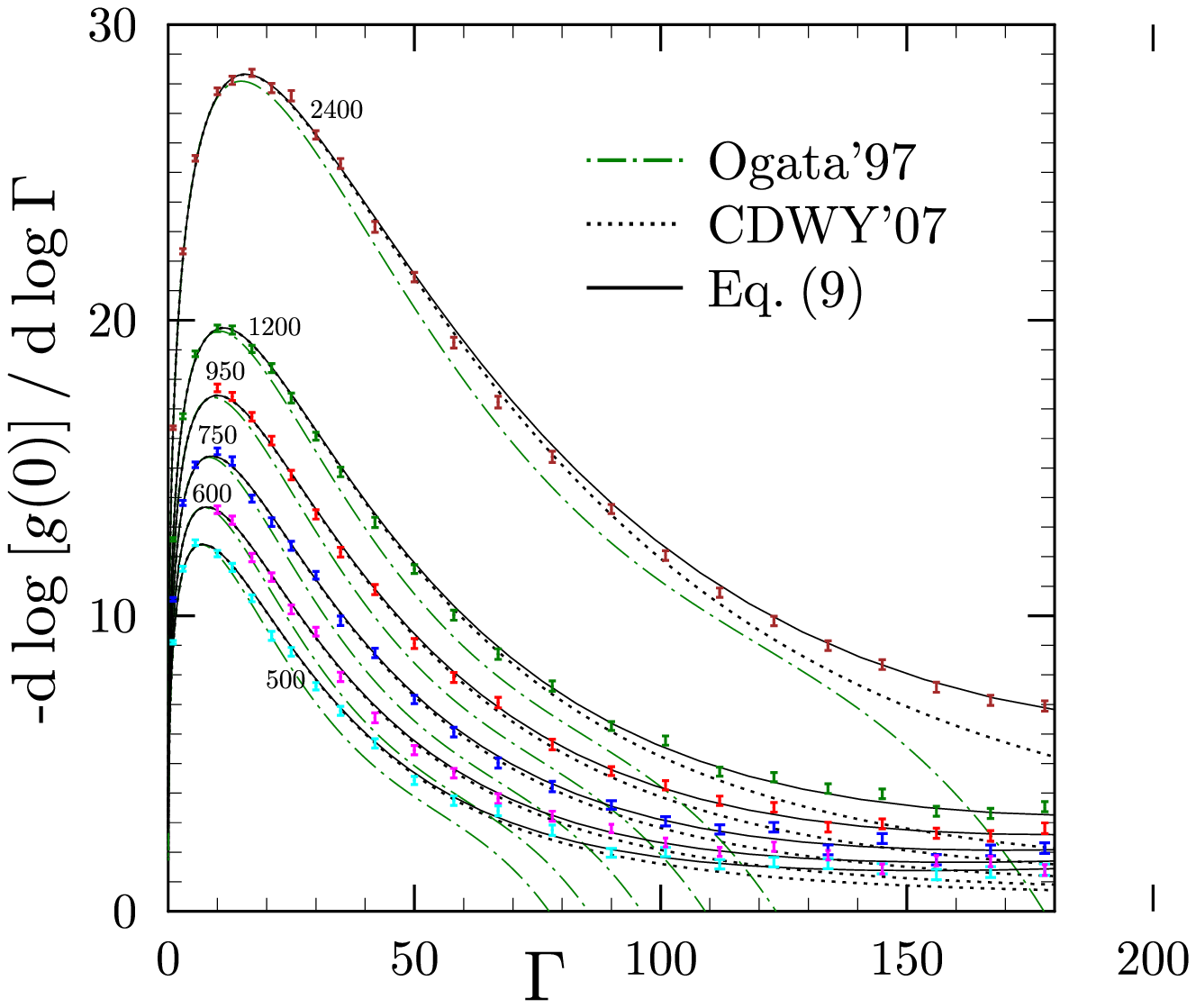} 
\end{center}                                                
\vspace{-0.4cm} 
\caption[]{Bars show the present PIMC results at the maximum number of 
beads. Colours highlight the data with the same $r_s$. Dotted, 
dot-dashed, and solid curves are the parametrisation of 
\citet{CDWY07}, \citet{O97}, and the fit, Eq.\ (\ref{RPIMC}), 
respectively, at $r_s$ values indicated near the curves.}                                             
\label{fit}
\end{figure}
%

%

In Fig.\ \ref{fit}, we show the final dataset. 
Calculations were performed at 6 values of $r_s$ and multiple values 
of $\Gamma$ in order to sample as best as practicable the deviation 
from the analytic parametrisation of \citet{CDWY07} plotted by the 
dotted curves. Different colours serve to highlight data groups 
sharing the same $r_s$. The solid curves provide an analytic fit to 
the PIMC data. They are obtained by adding a simple expression, 
$10^{-6} \Gamma^2 \sqrt{r_s}$, to the dotted curves.

This expression can be easily integrated over $\ln{\Gamma}$ to yield a 
correction to $g(0)$ of \citet{CDWY07}:
\begin{eqnarray}
    g_{\rm PIMC}(0) &=& g_{\rm CDWY}(0) F_{\rm PIMC}~,
\nonumber \\
    F_{\rm PIMC} &=& \exp{\left[- 5 \times 10^{-7} \Gamma^2 \sqrt{r_s} 
                     \right]}~.
\label{RPIMC}
\end{eqnarray}
In this case, $g_{\rm PIMC}(0)$ is our final result for the ion pair 
correlation function at zero separation.  
The respective modification of the reaction rates is depicted in 
Fig.\ \ref{rates}. The ratios of the reaction rates to those reported by 
\citet{CDWY07} are plotted as functions of $\Gamma$ for $r_s=500$, 600, 
750, 950, 1200 and 2400. The lower group of curves shows the 
present results while the upper group of curves displays the 
predictions of \citet{O97}. We conclude that in the physically relevant 
range of temperatures and densities, \citet{CDWY07} overestimate 
the reaction rates by no more than a factor of two whereas rates of 
\citet{O97} are overestimated by up to a factor of 30.

In the process of working on this paper we have 
calculated anew the energies of quantum liquid OCP in the density range 
$500 \leq r_s \leq 2400$ which turned out to be in a very good agreement
with the earlier computation and fit \citep[][]{B19,BY19}. This 
development will be reported in more detail elsewhere 
(Baiko \& Chugunov, submitted).

\section{Conclusion}
A first-principle method to calculate the ion pair correlation 
function and plasma screening factor in a Coulomb plasma is developed. 
The method is applied for an analysis of nuclear reaction rates in a 
practically relevant range of temperatures and densities typical for 
liquid layers of white dwarf cores and neutron star crusts. It is 
shown that among the earlier results on ion screening the most accurate 
are those of \citet*{CDWY07}. Nevertheless, at $\Gamma \gtrsim 100$  
a deviation of screening factors from analytic fits of these
authors is discovered which implies a reduction of the reaction rates
by up to a factor of two. A simple formula is proposed to fit the 
respective correction to the reaction rates valid for all $\Gamma$ and 
$r_s$ in the studied range. 

The developed method can be extended 
to the case of crystallised phase of stellar matter as 
well as to ionic mixtures of various elements which can be 
present simultaneously in stellar matter undergoing burning (e.g., 
carbon and helium). In these systems, currently existing theoretical 
uncertainty of the rates is higher than in the one-component liquid 
plasma considered here.

Another possible extension of the method could be to take into account 
the electron screening in a self-consistent manner. As long as the 
electron screening can be viewed as static, this reduces to a 
modification of the inter-ion potential. However, it is not 
straightforward, because such a potential should not be described by 
the simple Thomas-Fermi screening model \citep[][]{PC13}. A more 
realistic electron dielectric function and potential would have to be 
used which may significantly slow down the simulations. Curently, the 
electron screening of nuclear reactions is taken into account on the 
basis of the free energy difference in a classic ion plasma with 
{\it polarisable} background. 

A reliable description of nuclear fusion reactions in the liquid phase
of stellar matter is important for self-consistent modelling of various
astrophysical phenomena involving accretion onto a compact star or a 
merger of such stars. A local density increase results in an 
intensification of reactions in a matter element, an energy 
release, possible melting, and further intensification of reactions. 
Reconstructing the exact dynamics of this process is crucial for a 
realistic description of such spectacular transient phenomena as 
bursts, superbursts, and even Supernovae Ia.

\section*{Acknowledgements}
The author is sincerely grateful to D. G. Yakovlev for providing 
numerical data for Fig.\ \ref{strip}, reading the manuscript, and 
valuable comments.

\section*{Data Availability}
The data underlying this article will be shared on reasonable 
request to the corresponding author.

\end{document}